# Extreme Ultraviolet Time-Resolved Photoelectron Spectrometer with an Ultrathin Liquid Flat Jet

*Masafumi Koga,* [1, a)] *Do Hyung Kang,*[1] *Zachary N. Heim,*[1, b)] *Neal Haldar,*[1] *and Daniel M. Neumark*[1,2 c)]

[1]*Department of Chemistry, University of California, Berkeley, California 94720, USA*

[2]*Chemical Sciences Division, Lawrence Berkeley National Laboratory, Berkeley, California 94720, USA*

[a)] Current Address: Institute for Molecular Science, National Institutes of Natural Sciences, 38 Nishigo-Naka, Myodaiji, Okazaki 444-8585, Japan.

[b)] Current Address: Idaho National Laboratory, 1955 N. Fremont Ave., Idaho Falls, ID 83415, USA.

[c)] Author to whom correspondence should be addressed: dneumark@berkeley.edu

# Abstract

A setup for extreme-ultraviolet time-resolved photoelectron spectroscopy (XUV-TRPES) of liquids is described based on a gas-dynamic flat jet formed by a microfluidic chip device. In comparison to a cylindrical jet that has a typical diameter of 10-30 μm, the larger surface area of the flat jet with a width of ca. 300 μm allows for full overlap of the target with the pump and probe light beams. This results in an enhancement of photoelectrons emitted from the liquid, while simultaneously allowing smaller sample consumption compared with other flat jet techniques utilizing liquid collisions or converging slits. Femtosecond pulses of XUV light at a photon energy of 21.7 eV are prepared by high harmonic generation and a multilayer mirror that selects a single harmonic; the He gas used to form the gas-dynamic flat jet is transparent at this energy. Compared to a cylindrical jet, the photoelectron signal from the liquid is enhanced relative to that from the surrounding vapor jacket. Pump-probe spectra for aqueous thymine show notably higher signals for the flat vs cylindrical jet. Moreover, the time-dependent space-charge shift in UV pump/XUV probe experiments is smaller for the gas dynamic flat jet than for a cylindrical jet with the same flow rate, an effect that is accentuated at higher He backing pressures that yield a thinner jet. This reflects reduced multiphoton ionization of the solute by the UV pump pulse, the primary cause of the space charge shift, as the jet becomes thinner and reaches the thickness of a few tens of nm.

## 1. Introduction

Time-resolved photoelectron spectroscopy (TRPES) is versatile probe of ultrafast photophysical and photochemical processes that can be applied to gases, solids, and liquids.[1-6] The application of TRPES to volatile liquids such as water has historically presented a significant challenge owing to stringent vacuum requirements in photoelectron energy analyzers and the short mean free path of photoelectrons in liquids that can alter the observed photoelectron kinetic energies.[7, 8] The liquid microjet technique developed by Faubel and co-workers[9-11] successfully addressed this issue by providing a clean liquid surface for collision-free photoelectron detection. The probe photon energy of liquid jet TRPES has recently been extended to the extreme ultraviolet (XUV) region with a tabletop femtosecond light source using high-harmonic generation, thereby allowing universal ionization of valence electrons of pure liquids and solutes in microjets.[12-19] XUV-TRPES is therefore capable, in principle, of tracking the complete relaxation process of a given excited state from initial excitation to thermalization of the ground electronic state, and is particularly sensitive to non-adiabatic transitions since the vertical ionization energies of the electronic states associated with these transitions are distinct.[1, 20]

Nevertheless, using an XUV pulse to investigate solute photodynamics inherently poses a distinct set of challenges. XUV light is capable of ionizing water, whose vertical ionization energy is 11.33 eV,[21] in addition to the considerably more dilute solute. The pump-probe signal of a typical solute of interest is thus many orders of magnitude smaller than that from the water and can potentially be overwhelmed by photoelectrons originating from the solvent. In addition, the foci of the light beams used in these experiments are typically larger than a cylindrical jet diameter (<30 μm). Consequently, much of the XUV photon flux does not intersect the liquid, reducing the liquid photoionization signal at the expense of photoionizing the surrounding vapor and gas molecules. Furthermore, unless the UV pump pulse is sufficiently weak, there are time-dependent space charge effects induced by UV multiphoton ionization of water and solutes that can distort the time resolved spectra.[12, 13, 22]

In recent years, several groups have shown that flat jets of water and other volatile liquids can be incorporated into spectroscopy and scattering experiments requiring high vacuum.[15-17, 23-31] Flat jets can be over 1 mm wide and offer significantly better spatial overlap with incident light or molecular beams. Their thickness ranges from 10 nm to 7 μm, depending on the how the flat jet is formed and where it is probed.[31-33] The most prevalent approach to generating a flat jet is to utilize a colliding-jet configuration, whereby two cylindrical jets converge to expand radially in the direction perpendicular to the collision axis. The colliding jet has been employed in molecular beam scattering,[24-26] and multiple XUV and soft X-ray spectroscopies[17, 18, 27, 29-31]

The gas dynamic jet is an alternative way to form a flat sheet in vacuo: a high-pressure gas (typically He) is sourced from both sides of a cylindrical jet, which then acts to push and widen the jet, thereby forming a liquid sheet.[23] This jet can be operated at a liquid sample consumption rate as low as 0.1–0.2 mL/min, which is more than an order of magnitude smaller than the typical value of 1-10 mL/min in colliding jets. Moreover, Koralek[23] and De Angelis[28] demonstrated that a gas-dynamic jet operated with background He pressures above ~300 kPa forms an ultrathin liquid sheet with a thickness of ~10 nm. This feature extends the capability of X-ray absorption spectroscopy (XAS) of water solutions beyond the conventional water window.[28]

In this article, we describe a simple yet powerful XUV-TRPES setup based on a gas-dynamic flat jet. Solute molecules are photoexcited by an ultraviolet (UV) pump pulse and ionized by an XUV probe pulse. The XUV pulse is prepared using a 400 nm femtosecond pulse to drive high harmonic generation (HHG).[34] The 7th harmonic of the driving pulse at 21.7 eV is isolated through a series of metallic filters and a multilayer mirror. This isolation scheme obviates the necessity for a time-preserving monochromator, thereby simplifying the optical setup. The flat jet width of ~300 μm allows for full overlap of the pump-probe laser cross sections with the liquid, resulting in a larger liquid signal compared to a cylindrical jet. Simultaneously, the use of He gas, transparent at 21.7 eV, effectively sweeps away the water vapor jacket surrounding the surface, suppressing photoelectron signal from vapor photoionization. Furthermore, we find that the multiphoton photoelectron signal with the UV pump pulse diminishes as the He gas pressure increases and the jet thickness decreases, reducing time-dependent space charge effects on photoelectrons ejected from the liquid. This reduction is advantageous for measurements requiring high UV pump power and long time delays. Results from this instrument have been recently reported[15, 16]; here we describe all aspects of the apparatus in more detail.

## 2. Generation and Isolation of XUV pulse

The optical setup for UV pump XUV-TRPES is depicted in Fig. 1.[15] A femtosecond oscillator/regenerative amplifier (Coherent, Astrella) produces 7 mJ, 35 fs laser pulses with a central wavelength of 797 nm and a repetition rate of 1 kHz. The output is split by a set of beamsplitters that direct 1 mJ to a third harmonic generation (THG) setup consisting of a pair of *β*-barium borate (BBO) crystals to generate a UV pump pulse centered at 266 nm. A second 1 mJ pulse is delivered to an optical parametric amplifier (Coherent, TOPAS) that can generate tunable light from 230 nm to near-IR, providing an alternate pump pulse. The remaining 5 mJ is used to generate XUV light via HHG. The 797 nm pulse is frequency-doubled with a 200-μm thick BBO crystal. Any unconverted fundamental is removed by a 1-mm thick dichroic beamsplitter and the doubled light at 400 nm is introduced to a semi-infinite gas cell (SIGC) interfaced to the vacuum region.[35] For photoelectron spectroscopy, isolation of a single harmonic for the probe pulse (the 7th harmonic at 21.7 eV in this work) is crucial so that photoelectrons are created by a single photon energy. Using a 400 nm driver pulse yields twice the harmonic spacing (6.2 eV) as an 800 nm pulse, facilitating harmonic selection and removing the requirement for a time-compensating monochromator.[36-38] Additionally, the shorter driver wavelength results in substantially XUV higher flux owing to ~$\lambda^{-x}$ (x = 3–6 depending on the wavelength)[39-42] scaling and fewer harmonics by lowering the cut-off energy.[43]

The SIGC is a 60-cm long tube 1" in diameter with a 1-mm thick fused silica entrance window secured on a 1.33" conflat (CF) flange. At the cell exit, a 0.75" diameter, 250-$\mu$m thick stainless steel disk is coupled to a 10" CF 6-way cross chamber pumped by a 2000 L/s turbomolecular pump (Shimadzu, TMP-2003LM). Kr gas (99.9999%) is used as the gas medium for HHG; the SIGC pressure is monitored by a Baratron pressure gauge (MKS Instruments, 522C13TBE) and kept around 3 torr. The 400-nm, 900-nJ driving pulse is focused by a 1 m focal length lens to ca. 1 cm beyond the gas cell exit to maximize the harmonic flux. The strong photon flux at 400 nm drills a hole in the stainless-steel end cap, typically 200 μm in diameter, which serves as a pinhole that allows the harmonics to exit the gas cell. Due to slight variations in pointing, the pinhole eventually increases in size, necessitating replacement of the end cap every 1–2 months to maintain low pressure ($10^{-5}$ torr) in the differential pumping region. This region is connected by an 85 cm long, 3.8 cm diameter tube to the next chamber, a 10" CF 6-way cross pumped by a 440L/s turbomolecular pump (Leybold, TurboVac 450i). The long, narrow connecting tube ensures low pressure ($10^{-6} - 10^{-7}$ torr) in the downstream chambers and allows for sufficient divergence of the 400 nm laser beam to avoid burning optical elements.

In the chamber connected to the SIGC, the light propagates through a 200-nm thick Al filter backed by a Ni mesh (Lebow Company). This blocks the driving 400-nm laser and any harmonics below 15 eV, as shown in Fig. 2a. This chamber can also accommodate a custom-made spectrometer[44] where the harmonics after passing through an entrance slit are diffracted by a SiN$_x$ nanograting (LumArray, 100 nm periodicity) to be angularly separated by wavelength. The relative flux of the harmonics can be readily monitored by a detector comprising 2" chevron-stacked microchannel plates coupled to a phosphor screen.

The XUV light transmitted through the Al filter is then reflected by a toroidal focusing mirror at a grazing angle of 4º. This mirror has horizontal and vertical radii of curvature of 12323 and 94 mm, respectively, and re-focusses the XUV beam from the SIGC onto the sample at a distance of ca. 2 m. After the toroidal mirror, the XUV beam impinges on a multilayer (ML) mirror consisting of SiC and Mg layers at an incident angle of 45º, where the total layer thickness and layer thickness ratio are designed by the Center for X-Ray Optics (CXRO) at Lawrence Berkeley National Laboratory to optimize the reflection of the 7th harmonic of the 400 nm driving pulse (21.7 eV) while minimizing reflectivity of the 5th and 9th harmonics (15.5 eV, 27.9 eV). This mirror is held by a piezoelectric motor-actuated optical mount (Newport, 8821-AC-UHV), allowing for extremely fine spatial alignment of the XUV probe. The light finally passes through a 300-nm Sn filter before entering the sample interaction region.

Figure 2a shows the transmittance of the metallic filters and the reflectance of the ML mirror, obtained from the CXRO database,[45] and the total transmission of the setup is plotted in Fig. 2b. The reflectance of the toroidal mirror is not shown but should be constant over 15–40 eV. As shown in Fig. 2b, the XUV beamline successfully attenuates the transmission of the 9th harmonic by 6 orders of magnitude, yielding monochromatic 7th harmonic light at 21.7 eV. Since pump-probe signals of solutes are typically 5–6 orders of magnitude smaller than the solvent ionization signals, it is essential to keep the high eKE regime free of background, otherwise any contribution from the higher harmonics can overwhelm pump-probe signals of interest. For diagnostic purposes, the ML mirror can be replaced with a gold mirror that reflects multiple XUV harmonics. The chamber that houses the toroidal and multilayer mirrors is pumped by a 440 L/s turbomolecular pump (Leybold, TurboVac 450i).

The UV pump beam is coupled to the XUV beamline before the sample chamber by a custom made 25.4-mm annular concave mirror that has a 5 mm diameter aperture and a 2 m radius of curvature. The aperture transmits the XUV beam and is offset ca. 4 mm from the edge of the mirror. The UV pump light is coupled into vacuum through a 25.4-mm radius, 2 mm-thick CaF$_2$ window and guided to the remaining area of the annular concave mirror at 5º incident angle by a piezoelectric motor-controlled aluminum mirror. This configuration allows a minimal pump-probe crossing angle of 0.5º at the sample position. The focal position of the pump beam is precisely controlled by the translational stage mounting the concave mirror. This vacuum region is pumped by a 150 L/s turbo pump (Leybold, Turbovac 151) and is separated from the liquid jet chamber by a gate valve. The cross correlation of the XUV pulses and the third harmonic of the fundamental (~266 nm) is found to be 26 fs (1σ) and that with the TOPAS output (at 240 nm) is 35 fs by measuring the laser-assisted photoelectron (LAPE) spectrum in Ar gas.[46-48]

## 3. Liquid Jet Interaction Region and Photoelectron Spectrometer

Figure 3a shows an overview of the interaction region, including the flat jet mount assembly (top) and the magnetic bottle time-of-flight (MB ToF) spectrometer that consists of a magnetic cone on the right side of the figure and a ToF drift tube on the left. Photoelectrons enter the drift tube through a skimmer with an orifice diameter of 500 μm.

The liquid jet is prepared using a microfluidic chip device (Micronit, SKU: 11000957, Figure 3b) with three etched channels in borosilicate glass. The central channel with a diameter of 30 μm is supplied by one inlet, while the other inlet is connected to the two outer channels, each with a 50 μm diameter. The chip is operated in the gas-dynamic mode in which the cylindrical liquid jet from the central channel is flattened by the He gas jets from the two outer channels. Operation in the colliding jet mode is described in detail elsewhere.[24] The chip is mounted on a custom-made chip holder with two PEEK inlets for both liquid and gas jets. The chip holder is attached to an optical post mounted on a piezoelectric motor-controlled rotational stage (Thorlabs, ELL18K); this in turn is mounted to a three-axis translational stage with 1" travel linear piezoelectric actuators (Newport, 8301NF). The combination of stages allows precise 3D manipulation of the jet angle and position. The flat jet position with respect to the skimmer orifice is manipulated while monitoring the liquid water photoelectron signal and is typically around 1.0 mm from the skimmer. The flat jet angle $\theta$ (Fig. 3c) is defined as the liquid surface normal with respect to the XUV beam axis and typically fixed at 60° since this results in maximal signal (see Section 5-1).

Liquid sample is delivered to the chip by an HPLC pump (Shimadzu, LC-40i) through PEEK tubing after a set of in-line frits made of either PEEK or stainless steel. After crossing the laser interaction region, the liquid jet is frozen out on a liquid nitrogen-cooled trap located about 30 cm below the nozzle. The trap is 1 ft long with a diameter of 2 1/2 inch and can capture ca. 900 mL of water.

The entire liquid-jet interaction region and MB ToF spectrometer assembly is housed within a series of two chambers and is mounted on a custom-built xy translation frame that has 2" movement in both directions to enable alignment with the annular mirror assembly. The trap chamber containing the liquid jet mount is directly connected to the annular mirror chamber through a 2.75" bellows with a gate valve. A 10" CF door is attached on the other end of the chamber for easy access to the jet assembly. The door includes two 2" windows; a centered window for the laser exit and an off-axis window for viewing the liquid jet using a USB camera (Imaging Source, DFK 37AUX226) attached to a macro video lens (Navitar, Zoom 7000). The laser exit window is also used to illuminate the flat jet with a desk lamp so that one can view the jet while running in vacuum.

The chamber holding the liquid jet is attached to a 10" CF 4-way cross pumped by two turbomolecular pumps (Pfeiffer Vacuum, ATH1603M and Leybold, Turbovac 450i) with a total pumping speed of 1790 L/s and a custom-made cryogenic cold trap. The liquid nitrogen-filled cold trap, whose total pumping surface area is about 1300 cm$^2$, provides a theoretical pumping speed of 19,000 L/s for H$_2$O. Typical operating pressures are $3\times10^{-5} - 9\times10^{-4}$ torr, depending on the He gas pressure.

The magnetic bottle ToF spectrometer has been described in detail elsewhere,[49] but several aspects of its design required optimization for flat jet operation with an XUV probe pulse. The inhomogeneous magnetic field in the interaction region is produced by a stack of four neodymium disk magnets with a total height of 1" and a soft iron cone (Ed Fagan, HyperCo 50A) tapered at a 45º apex angle is designed to focus the magnetic flux density on the tip. The magnetic cone stack is mounted on a XYZ stage attached to a customized rotation stage. The alignment of each axis is achieved under vacuum with a piezoelectric motor (Newport, 8301NF). The magnetic field strength is determined by numerical simulations by the Finite Element Method Magnetics (FEMM) software; it is approximately at 1.1 T at the tip of the cone and 0.3 T at the point of the laser interaction region located around 5 mm from the cone tip.

The photoelectron detector region is isolated from the laser interaction chamber by the aforementioned skimmer mounted on a differential pumping sheath. The ToF tube is 66-cm long. It is mounted within a solenoid consisting of 14 AWG copper wire coiled at 10 turns per inch and operated at a current of 2A that generates a weak homogeneous field of 8G within the flight tube. Photoelectrons are detected by a chevron-mounted microchannel plate stack (Photonis Scientific, Inc.) with a plate diameter of 25 mm coupled to a phosphor screen used as a signal voltage readout. The phosphor screen is also useful in checking the alignment of the jet and permanent magnets with respect to drift tube axis, since the electron signal appears as a small spot in the center of the screen when the alignment is correct. The entire detector region is pumped out by a set of turbomolecular pumps (Varian, Turbo-V 550; Leybold, Turbovac 1000C and Turbovac 150) with a total pumping speed of 1800 L/s for $N_2$. The vacuum pressure in the detector region is typically ca. $6\times10^{-6}$ torr during operation.

## 4. Characterization of the Monochromatic XUV Pulse

The XUV spectrum is characterized by measuring photoelectron spectrum of argon, for which ionization potential of 15.76 eV. Spectra obtained with the gold mirror only and the multilayer mirror/Sn filter are shown in Fig 4. With the gold mirror, the photoelectron signal comprises multiple peaks spaced by 6.2 eV. Even though the 7th harmonic is strongest, the 9th and 11th harmonics at 27.0 and 34.1 eV can be problematic since solvent ionization signal from these harmonics can overwhelm the intrinsic UV pump-XUV probe signal of interest from the 7th harmonic. The ML mirror and Sn filter effectively eliminate the 9th and 11th harmonics, providing a background-free electron detection window for the pump-probe experiments. Although the transmission of the Sn filter is only 10% at 21.7 eV, the elimination of background from the higher energy harmonics is well worth it. A typical number of XUV photons with the ML mirror and the Sn filter measured by a photodiode (OptoDiode Corp, AXUV100G) is $10^9$ photons/pulse.

The measured width of the Ar peak with the ML mirror and Sn filter is ~0.4 eV. There are three contributions to this: the bandwidth of the XUV light, the energy resolution of the ToF spectrometer, and the unresolved $Ar^+$ spin-orbit splitting of 0.18 eV. The timing precision of the ToF spectrometer is determined by the ~3 ns pulse width of detected electron bunches, corresponding to an energy spread of 0.07 eV for photoelectrons with eKE = 6 eV. Based on the latter two contributions to the width of the Ar peak, we estimate the bandwidth of the XUV pulse to be ~0.35 eV. Note that the minor peak around 9.7 eV eKE is due to the ionization of contaminant oxygen for which the ionization potential is 12.07 eV.

## 5. Specification of the Flat Jet PES

### 5-1. Sheet Size and Angle Dependence

Figure 5 shows a series of pictures of the flat jet under various He gas pressures in which the jet is illuminated with broadband visible light. The scale bar shown in Fig. 5a is derived from the known chip width, 800 μm. At a gas pressure of 200 kPa (Fig. 5a) we observe a small primary sheet with maximum width of ~100 μm, followed by downstream secondary and tertiary sheets. The primary sheet becomes larger with increasing He pressure and reaches a maximum width of ca. 300 $\mu$m at 450 kPa (Fig. 5d). Reflection of white light on the sheets reveals radial interference fringes emanating from the chip orifice that are most recognizable in Fig. 5b. The fringe pattern in Fig. 5b persists for ca. 100 μm from the chip orifice and fades downstream until the re-convergence point, where the primary sheet ends. At higher pressures (5c and 5d), the fringes become more closely spaced and persist for a shorter distance, with all reflectance disappearing about 300 μm downstream from the jet orifice. Koralek and coworkers[23] ascribed the interference patterns to rapidly changing spectral reflectance due to a decrease in sheet thickness from 1 μm to ~200 nm as one moves downstream from the jet orifice, with the fringes disappearing when the jet thickness drops below 100 nm. The pictures in Figure 5 thus show that as the He backing pressure is raised, the jet becomes thinner and wider. The disappearance of reflected light in the lower half of the first jet sheet in Figures 5c and 5d indicates that the jet thickness is less than 100 nm, and likely approaches the value of 20 nm extracted by Koralek from spatially resolved infrared absorption measurements.

Photoelectron spectra of a water (NaCl 25 mM) flat jet at various input gas pressures are shown in Fig. 6a. The laser angle of incidence $\theta$ (Fig. 3c) is set to 60°. Optimization of the jet position is required for each He pressure, in particular along the z-axis (height) so that the XUV pulse always intersects the center of the leaf. The broad peak at 11 eV is from ionization out of $1b_{1(liq)}$, the $1b_1$ orbital of liquid $H_2O$, the two narrower peaks around 12 eV are from $1b_{1(gas)}$, and the signals at higher eBE are from a combination of liquid and gas phase ionization.[50-52] Signals above 16 eV are dominated by inelastic background.[53] The overall signal, particularly that originating from the liquid phase of water, becomes more pronounced at the higher gas pressures. The upper panel in Fig. 6a shows the difference between the spectra at 440 and 220 kPa. While the positive feature around 11 and 13 eV, corresponding to the $1b_{1(liq)}$ and $3a_1$ peaks, is present, strong depletion is observed at the $1b_{1(gas)}$ regime, indicating a reduced contribution from gaseous $H_2O$ at higher He pressure. This result shows that the He gas sweeps away the $H_2O$ vapor sheath, enabling selective enhancement of the target liquid signal while suppressing the gas phase background.

Figure 6b depicts the signal intensities of the $1b_{1(liq)}$ at 11.0 eV and $1b_{1(gas)}$ at 12.1 eV as a function of the He pressure; the latter spectrum is obtained by subtracting the pure liquid spectrum[50] (dashed blue line in Fig. 6a). The liquid signal increases as the helium gas pressure rises to approximately 300 kPa, above which it remains almost constant. This suggests that above 300 kPa, the flat jet is sufficiently wide to encompass the entire XUV focal spot, and that further expansion of the sheet does not contribute to the signal increase. However, the gas peak intensity exhibits a monotonic decrease even in the regime where the liquid peak intensity saturates, again indicating that the He gas jets sweep the water vapor away from the surface and suppress the water gas peak.

The rotational mount accommodating the chip enables one to optimize the jet angle in situ. Figure 7a shows the peak intensity of water $1b_1$ photoelectrons as a function of the laser angle of incidence $\theta$. The photoelectron signals from both the liquid and gas phases increase with $\theta$ up to 60º, followed by a drop in intensity of the liquid signal while the gas signal continues to increase. The liquid signal intensity is greater than that of the gas peak for angles between 20º and 70º. Figure 7b shows the angle-dependent liquid to gas intensity ratio, found to be maximized at $\theta = 60°$. A similar plot previously obtained with a colliding flat jet is plotted in the same figure.[27] Comparison of the two data sets shows that the liquid:gas signal ratio is consistently higher in the gas-dynamic mode of operation, further pointing to the role of the He gas in reducing photoelectron signal from the vapor sheath surrounding the flat jet.

**5-2. Comparison of pump-probe signals for flat and cylindrical jets**

The dependence of the pump-probe signals on the type of jet is investigated using aqueous solutions of thymine (Thy), a system that Miura et al have previously studied in detail with XUV-TRPES using a cylindrical jet.[14] Figure 8 shows the baseline-subtracted TRPE spectra of Thy in a flat jet (310 kPa, Fig. 8a) and a cylindrical jet (30 $\mu$m, Fig. 8b). These data were collected using identical excitation pulse intensities and with 100k shots per delay time, corresponding to an hour of acquisition time with 40 delay points. The spectra are not corrected for the time-dependent space charge shift (Section 5-3). Both TRPE spectra exhibit a strong feature above 5 eV that only appears at the time origin. This signal is attributed to LAPE of the liquid water, as previously demonstrated.[15, 54] The pump-probe signal at the shortest delays, which represents the $\pi\pi^*$ excited state of Thy, is observed out to 3 eV eBE and decays within 500 fs, followed by a long-lived signal that persists for several ps (not shown). While the time evolution of the two spectra is similar, the pump-probe signal is approximately 1.5 times stronger in the flat jet than that in the cylindrical jet. It is noteworthy that the signal above 6 eV and beyond 500 fs turns negative with the cylindrical jet because the time-dependent space-charge effect is larger in the cylindrical jet, as described in greater detail in Section 5-3.

Figure 9 shows the time evolutions of the LAPE-subtracted photoelectron signal intensity integrated over 2.5–6 eV for the flat and cylindrical jets the two types of jet. The time evolution of the flat jet signal is in good agreement with that found by Miura et al out to 20 ps. While the FJ data point at 33 fs is off from the others due to the residual of the LAPE subtraction, it can be adequately reproduced by a double exponential function convoluted with the gaussian instrumental response function using time constants of 0.35 and 2.55 ps. The first time constant is consistent with the prior TRPES studies[19, 55] which ascribed the decay to the $\pi\pi^*$ state. The dynamics on the picosecond timescale, characterized by the second time constant of 2.55 ps and the offset (>20 ps) component are in line with the reported $n\pi^*$ lifetime and the long-lived $^3\pi\pi^*$ state, respectively.[14] In contrast, the time evolution of the cylindrical jet is difficult to discern beyond 1 ps because of the low signal-to-noise ratio and the time-dependent space charge shift.

**5-3. Space-Charge Shift**

One of the more problematic aspects of UV/XUV liquid jet TRPES experiments is that the spectra exhibit a time-dependent space-charge (SC) shift that increases with pump-probe delay. As discussed in depth by Hummert,[12, 56] this effect arises owing to multiphoton ionization (MPI) of the solute molecules in the jet by the UV pump pulse, which is typically resonant with a strong electronic transition in the solute, and the interaction of the positively charged jet created by the pump pulse with photoelectrons subsequently created by the XUV probe pulse.

Figure 10a shows an example of how the SC effect changes the photoelectron spectrum of a buffered 15 mM aqueous solution of thymine. The spectrum at a long positive time delay (40 ps) shows that the $1b_{1(liq)}$ peak occurs at higher eBE than at a negative time delay of -300 fs, while the $1b_{1(gas)}$ peak is unchanged. This shift arises from deceleration of the probe electron packet traveling in the electrostatic potential formed between the positive charge localized within the liquid jet and the expanding negative charge cloud created by MPI. The resulting shift toward lower eKE for the probe-generated photoelectrons increases with pump-probe delay time, as it takes longer for these photoelectrons to pass through the cloud of slower electrons formed by the pump pulse. For a cylindrical jet, Hummert and coworkers[12, 56] modeled this time-dependent shift approximating the pump laser-induced charge distributions as spherical shells, where the electron packet expands radially and the positive ions on the jet are static. While this model is not directly applicable to a flat jet owing to its lower symmetry, the underlying phenomenon is the same.

We next explore the SC shift as the jet thickness is varied. To retrieve the delay-dependent SC shift from pump-probe data, the water $1b_{1(liq)}$ and $1b_{1(gas)}$ peaks are fitted using double Gaussian functions, which yield the time evolution of the kinetic energy of the liquid water peak. An example of the fitting result is shown as dashed lines in Fig. 10a, in which the double Gaussian functions, one each for $1b_{1(gas)}$ and $1b_{1(liq)}$, can reproduce the spectra in the low eBE regime out to 40 ps. In the fitting procedure, the gas peak position and the liquid peak intensity are fixed. As shown in Fig. 10b, the SC shift increases at all He pressures with increasing pump-probe delay. However, the magnitude of the SC shirt clearly drops as the He pressure is raised, most notably from 140 to 270 kPa. For example, the shift between zero and 10 ps drops from 0.5 eV at 140 kPa to 0.1 eV at 510 kPa. Hence, as the jet becomes flatter and thinner, the time-dependent SC shift drops.

The reduction of the SC effect by the gas dynamic jet is of considerable interest since the elucidation of long-time dynamics is often an important goal in LJ-TRPES experiments, as shown in the work by Miura et al[14] for aqueous Thy solutions; the extraction of these dynamics is facilitated if the SC shift is reduced. While the SC shift can be reduced by lowering the UV pulse power, this results in less pump-probe single. In contrast, using a thin liquid jet in the gas-dynamic mode offers a relatively painless means of mitigating time-depending SC effects.

We explore the origin of the SC dependence on jet thickness by measuring pump-only photoelectron spectra of a 15 mM Thy solution at different He pressures. Figure 11a shows the results at 266 nm (4.66 eV) and 300 nJ per pulse. The peak position at ~1.4 eV eKE corresponds to (resonant) two-photon ionization of the ground state of aqueous thymine (VIE = 7.9 eV).[57] The overall MPI signal intensity decreases as the He pressure is increased. Moreover, the spectra at lower He pressures exhibit high-eKE tails extending to ~8 eV at 140 kPa, attributed to an internal SC effect in which the large number of photoelectrons broadens the UV photoelectron packet. In any case, at higher He pressures, fewer photoelectrons are produced, as shown in Fig. 11b where the integrated MPI photoelectron signal is plotted vs He pressure, and this means that charging of the jet by the UV pulse is also lower at higher He pressures. In Fig. 11c, the SC shift ΔE in a pump-probe experiment (at 12 ps) on the Thy solution using the same UV conditions is plotted vs the integrated pump-only photoelectron signal, which is proportional to the charges created by the pump pulse. The near linear dependence shows that the SC shift originates from charging of the jet by the UV pump pulse as was also expected in the cylindrical jet.[56] Hence, the reduction in UV-induced charging for a thinner flat jet leads to a smaller SC shift.

One must now consider why the multiphoton signal is less for a thinner jet. Since the path length of a thicker jet is longer, more MPI electrons are created within the thicker jet, i.e. away from the surface of the jet. However, for the jet to charge up, the electrons have to escape from the jet. The extent to which they can do so depends on the escape depth of electrons in liquid water, generally parameterized as the electron attenuation length (EAL), the length over which the photoelectron signal at its initial kinetic energy to be reduced to 1/e.[58-60] The EAL is only 1–3 nm for eKEs in the range of 10–20 eV, i.e. from the XUV pulse, because of high cross sections for electronic inelastic scattering. Below 10 eV, electron attenuation is dominated by elastic scattering and by intra- and intermolecular vibrational inelastic scattering.[50]

Experimental EAL values at this lower kinetic energy are uncertain but are generally believed to increase as the eKE drops from 10 eV. Some laboratories have reported inelastic mean free paths of ca. 5 nm for eKEs around 1–4 eV.[7, 8, 61] However, Fielding and coworkers[62, 63] have reported in their UV-PES studies coupled with Monte-Carlo simulations that photoelectron signal loss for 0.7–2 eV is negligibly small when electrons are generated within 25 nm of the microjet surface, suggesting that low eKE photoelectrons can escape from solutions through multiple scattering processes. One can thus have a situation where a non-negligible fraction of MPI events lead to net charging of the jet, whereas electrons created by XUV light that far from the interface will be attenuate to the extent that they will not contribute significantly to the photoelectron signal. What this means is that for a flat jet that is, say, only 20 nm thick, there will be less charging of the jet from MPI than for a thicker jet, hence a smaller space charge shift. Interestingly, MPI keeps decreasing while increasing the He gas pressure above 350 kPa, at which point the jet thickness is already expected to approach 10 nm for >200 µm from the jet nozzle according to the previous study on similar injection systems.[28] While this explanation needs to be made more quantitative, the correlation of the reduced SC with the very thin gas dynamic jet appears to be quite robust and represents a perhaps unexpected advantage in comparison to a cylindrical jet.

**Conclusions**

We described the development and characterization of XUV-TRPES setup that employs a gas dynamic flat jet. The XUV probe light is generated by HHG with a 400-nm driving pulse in a semi-infinite gas cell and is subsequently filtered by a multilayer mirror and a series of metallic filters, resulting in a monochromatic light source at 21.7 eV in the simple optical layout without monochromators. While this XUV energy allows full ionization of species of interest in solutions, the He gas used to form the gas-dynamics flat jet is transparent at this energy. The size of the flat jet generated by the microfluidic chip device can readily be controlled by the He gas pressure; above 300 kPa it is large enough to cover the whole XUV focal spot while suppressing the ionization of the gaseous water peak. Consequently, the pump-probe signal of thymine in the flat jet is notably stronger than that in a 30-µm cylindrical jet. The thickness of the flat jet is also controlled by the He gas pressure. Operation at higher pressures, where the jet is thinner, substantially mitigates the time-dependent space charge shift of the pump-probe signals at long delay times, a persistent problem in UV/XUV TRPES of liquid jets. We show that the photoelectron signal from UV-induced resonant multiphoton ionization of solute is lower for thinner jets, reducing charging of the jet and the consequent space-charge shift.

**Acknowledgments**

This research is supported by the National Science Foundation Division of Chemistry under Grant No. CHE-2422769. DHK acknowledges support from the Army Research Office under Grant No. W911NF-23-1-0003. Additional support was provided by CALSOLV, the center for solvation studies at the University of California, Berkeley. M.K. also acknowledges support from the Japan Society for the Promotion of Science (JSPS) Overseas Research Fellowships. D.H.K. acknowledges an award from Basic Science Research Program through the National Research Foundation of Korea (NRF) funded by the Ministry of Education with Grant No. RS-2023-00241698. The authors thank Walt Yang for his help in implementing the flat jet.

**Conflicts of interest**

There are no conflicts of interest to declare.

**Figures**

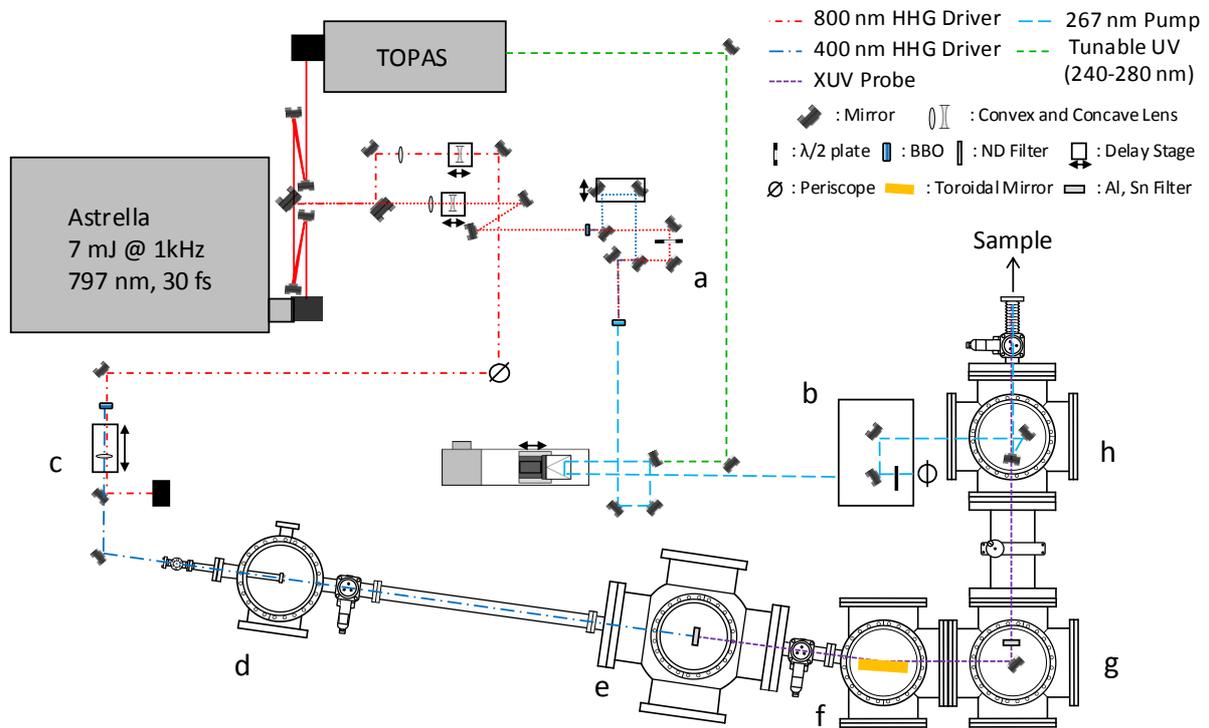

**Figure 1.** Schematic diagram of optical layout for UV pump/ XUV (21.7 eV) probe experiment. The optical path for the tunable UV (green) is illustrated only up to the 267-nm mirror in front of the delay stage, from where it follows the same path as the 267 nm. The pump UV pulse is generated by the TOPAS or (a) BBO-based THG, and coupled to the vacuum through (b) breadboard above table and periscope. The 400-nm HHG driver is prepared SHG (c) with the focusing lens. The vacuum region consists of (d) SIGC and differential pumping regime, (e) primary filtering and beam analyzer chamber, (f) toroidal mirror, (g) ML and Sn filter, (h) annular mirror chamber where the pump beam merges, and the sample detection chamber (not shown).

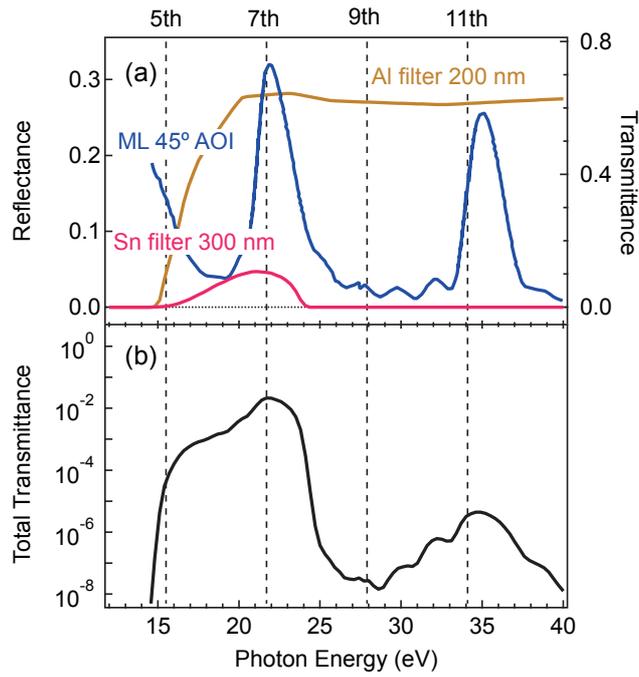

**Figure 2.** (a) XUV transmittance of Al filter (200 nm, yellow), Sn filter (300 nm, red) and reflectance of the multilayer mirror at an incident angle of 45º (blue). (b) Total transmittance of the XUV beamline with optics shown in (a).

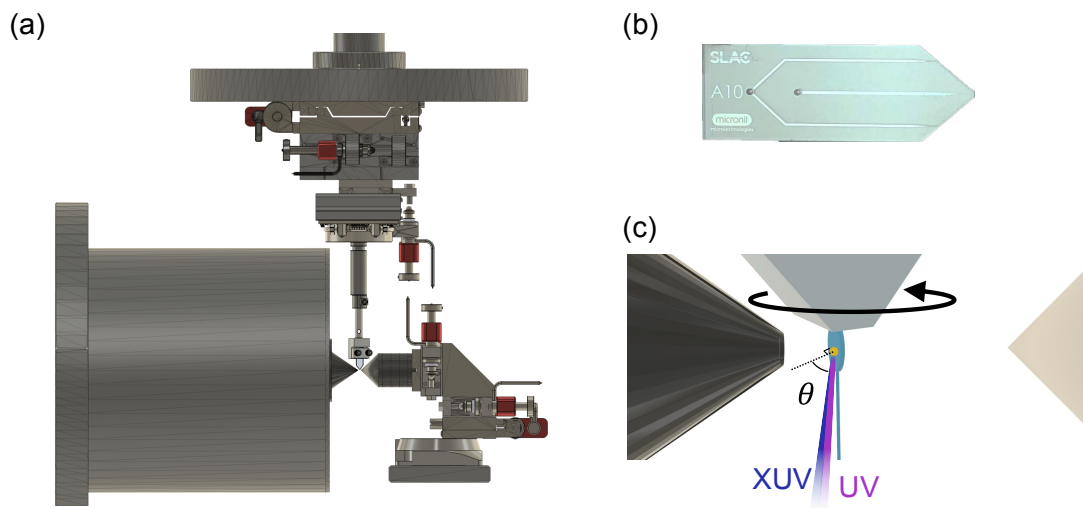

**Figure 3.** (a) Schematic of the interaction region coupled and the magnetic-bottle time-of-flight spectrometer. (b) Microfluidic chip device (Micronit) used in the experiment. (c) Magnified view of the interaction region showing the UV and XUV beams.

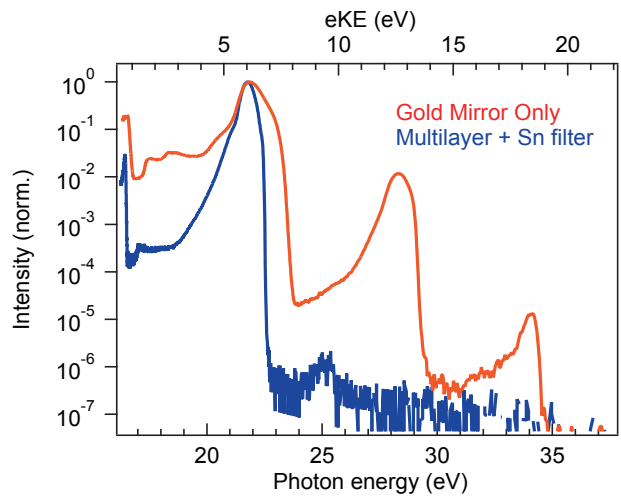

**Figure 4.** Photoelectron spectra of Ar obtained with the gold mirror (red) and with the ML mirror and Sn filter (blue).

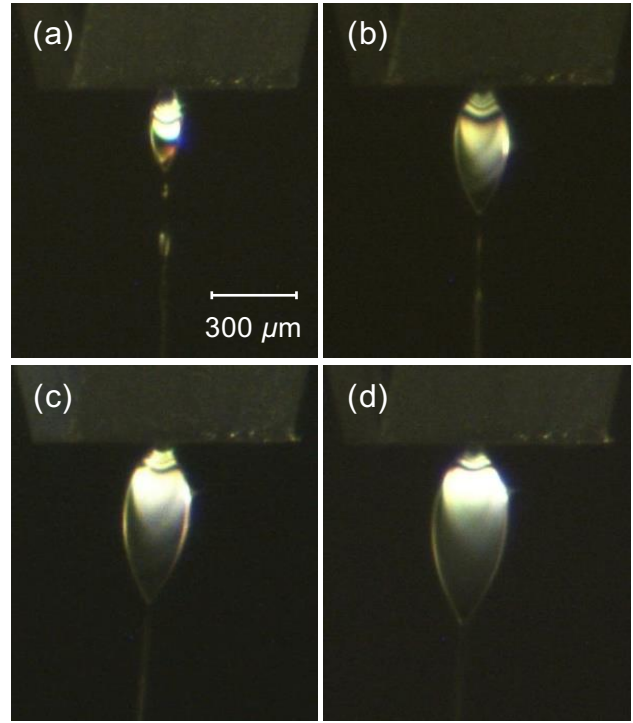

**Figure 5.** Gas-dynamic flat sheet of 25 mM aqueous solution of NaCl at input He gas pressures of (a) 200 kPa, (b) 300 kPa, (c) 400 kPa, and (d) 450 kPa. The solution flow rate is fixed at 0.2 mL/min. The CMOS camera (Imaging Source, DFG 37AUX226) is placed at an incident angle of 20º. The scale bar shown in (a) represents the scale parallel to the jet surface.

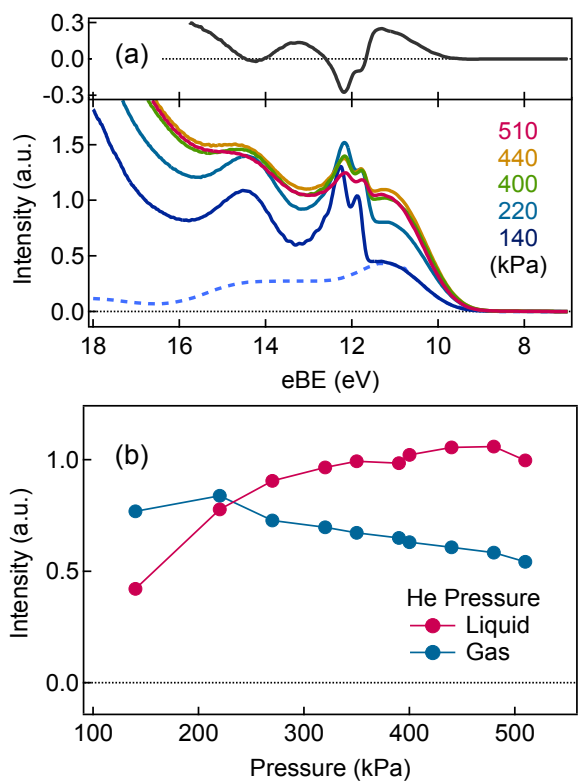

**Figure 6.** (a) XUV one-photon photoelectron spectra of water (Thy 15 mM in Trizma buffer 2 mM, pH=8, NaCl 25 mM) flat jet with various He gas pressure. The laser incident angle with respect to the jet surface is fixed at 60º. The difference spectrum between results at 440 and 220 kPa is shown in the upper panel. The dashed line represents the liquid water PES at hν = 25 eV reproduced from ref. 50, obtained by separating the gas phase spectrum by electrically biasing a cylindrical jet. (b) Photoelectron signal intensity of the liquid and gas peaks at 11.0 and 12.1 eV, respectively, as a function of He gas pressure. The intensity of the gas phase peak is obtained by subtracting the overlapping liquid-only signal (dashed curve in (a)).

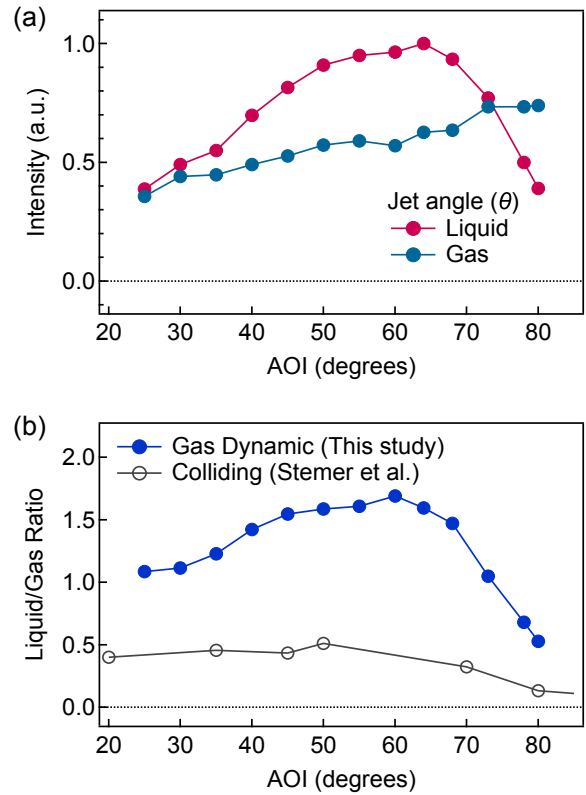

**Figure 7.** (a)Photoelectron intensity of the gas (blue, 12.1 eV) and liquid (red, 11.0 eV) peaks as a function of the jet angle ($\theta$: laser incident angle). He pressure is fixed at 400 kPa. The gas intensity is obtained after subtracting the liquid signal overlap. (b) Liquid to gas peak ratio as a function of the jet angle (blue). The ratio for a colliding jet obtained with oxygen 1s photoelectron spectroscopy (h$\nu$ = 640 eV) is also shown as gray open circles. Data reproduced from ref 27.

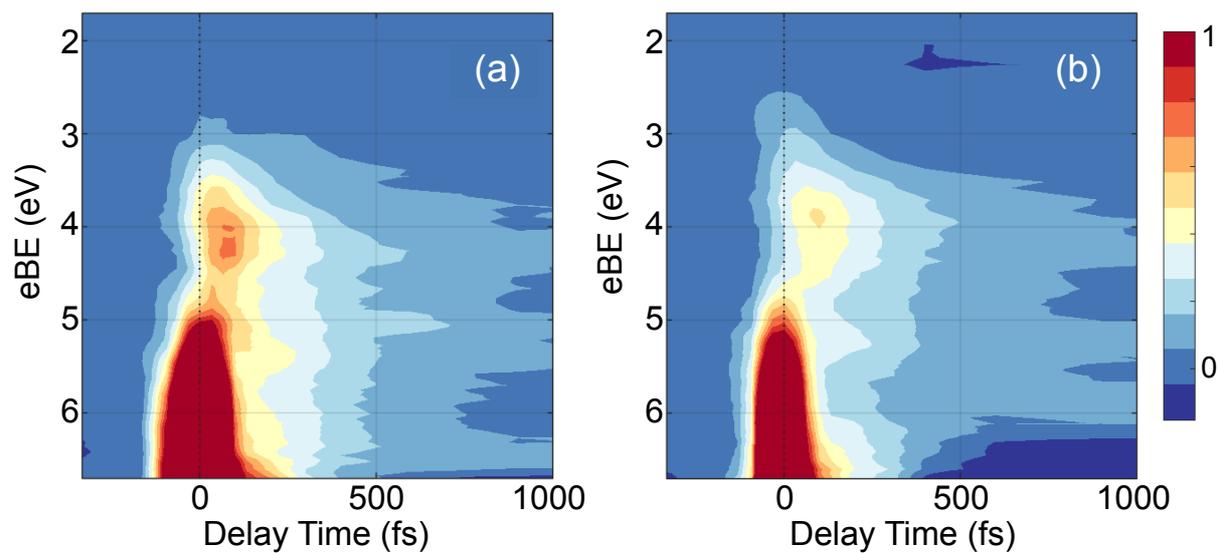

**Figure 8.** TRPE spectra of 15 mM Thy in buffered aqueous solution obtained with (a) the flat jet (He gas 310 kPa), and (b) the 30-μm cylindrical jet using 250 nJ pump-pulse excitation.

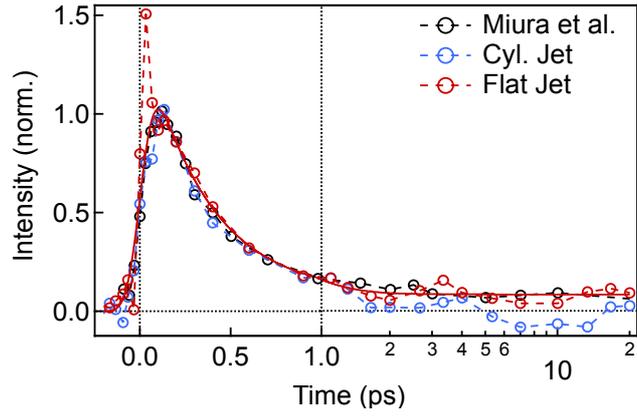

**Figure 9.** Time evolution of the LAPE-subtracted photoelectron signals obtained with the cylindrical jet (blue), and the flat jet (red), integrated over 2.5-6 eV. The red solid line shows the curve fitting results for the flat jet. The black circles represent the photoelectron signal reported in ref 14 obtained under the same integration range

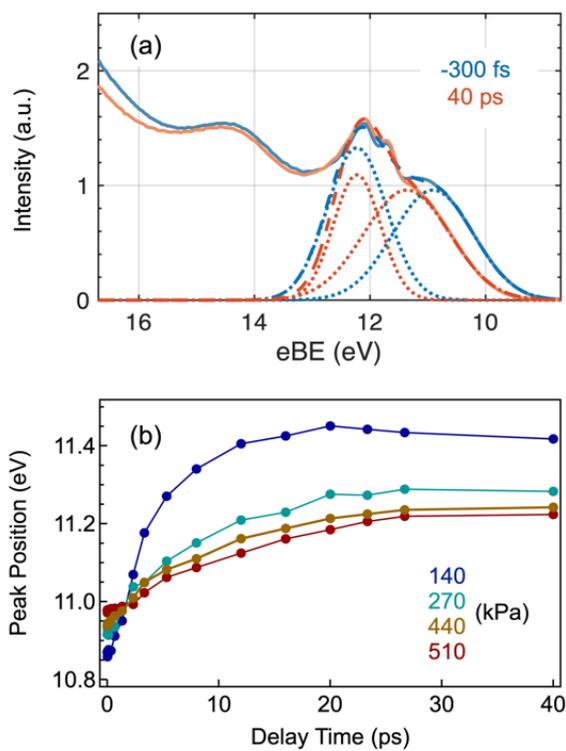

**Figure 10.** (a) Example of double-Gaussian fitting of the photoelectron spectra from the flat jet (15 mM Thy in Trizma buffer 2 mM, pH=8, NaCl 25 mM) at 320 kPa He pressure at negative (-300 fs, blue) and positive (40 ps, red) time delays. (b) Time evolution of the $1b_{1(liq)}$ peak positions (in eBE) at different He gas pressure obtained from the double-Gaussian fit, showing a reduced time-dependent space charge shift at higher He pressure.

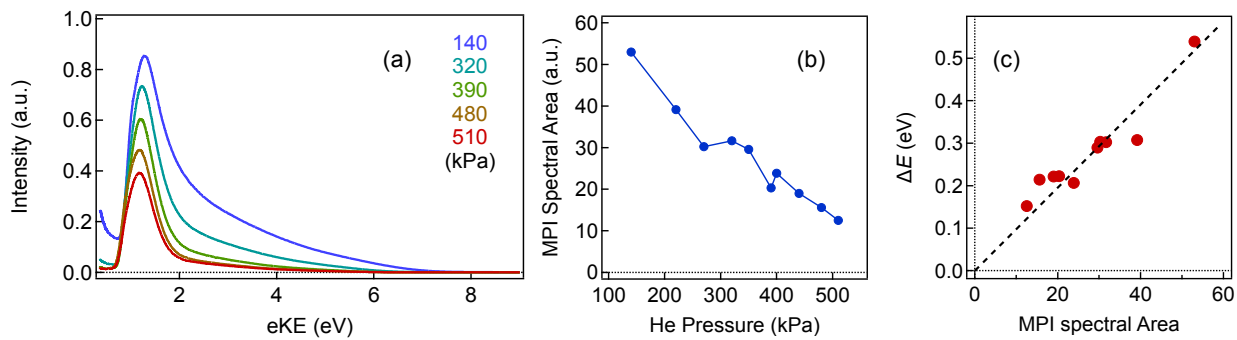

**Figure 11.** (a) Photoelectron spectra of Thy solution (15 mM, Trizma 2 mM, pH=8 NaCl 25 mM) with various He pressure under 266-nm pulse irradiation (300 nJ/pulse). (b) UV photoelectron signal area as a function of the He pressure. (c) SC amplitude (ΔE) at 12 ps depending on the MPI spectral area.